\documentclass[preprint,showpacs,preprintnumbers,amsmath,amssymb]{revtex4}

\usepackage{graphicx}
\usepackage{dcolumn}
\usepackage{bm}

\newcommand{\be}{\begin{equation}}  
\newcommand{\ee}{\end{equation}}  
\newcommand{\bear}{\begin{eqnarray}}  
\newcommand{\eear}{\end{eqnarray}}  
\newcommand{\ba}{\begin{array}}  
\newcommand{\ea}{\end{array}}

\newskip\humongous \humongous=0pt plus 1000pt minus 1000pt

\newif\ifdtup

\def\oldreffmt#1{\rlap{[#1]} \hbox to 2\parindent{}}

\def\figfmt#1{\rlap{Figure {#1}} \hbox to 1in{}}  
  
\def\ie{\hbox{\it i.e.}{}}	  
\def\eg{\hbox{\it e.g.}{}}

\def\bra#1{\left\langle #1\right|}  
\def\ket#1{\left| #1\right\rangle}  
\def\VEV#1{\left\langle #1\right\rangle}

\def\slash#1{#1\!\!\!/\!\,\,}  
\def\beq{\begin{equation}}  
\def\eeq{\end{equation}}  
\def\bea{\begin{eqnarray}}  
\def\eea{\end{eqnarray}}  
\def\half{\frac{1}{2}}  
  
\def\bq{\begin{quote}}  
\def\eq{\end{quote}}

\def\half{\frac{1}{2}}       
  
\def \gta {\mathrel{\vcenter  
     {\hbox{$>$}\nointerlineskip\hbox{$\sim$}}}}   

\relax  

\newdimen\tdim  
\tdim=\unitlength  
\def\bar{\overline}

\begin{document}

\preprint{FERMILAB-PUB-12-589-T}

\vspace{2.0cm}

\title{``SUPER''--DILATATION SYMMETRY \\
OF THE TOP-HIGGS SYSTEM}

\author{CHRISTOPHER T. HILL}

\affiliation{
 {{Fermi National Accelerator Laboratory}}\\
{{\it P.O. Box 500, Batavia, Illinois 60510, USA}; 
}
}%

\date{\today}

\email{ hill@fnal.gov }

\begin{abstract}
The top-Higgs system, consisting of top quark ({\em LH} doublet, {\em RH} singlet) 
and Higgs boson kinetic
terms, with gauge fields set to zero, has an {\em exact} (modulo
total divergences) symmetry where 
both fermion and Higgs fields are 
shifted and mixed in a supersymmetric fashion.
The full Higgs-Yukawa interaction and Higgs-potential, including
additional  $\sim 1/\Lambda^2$ NJL-like interactions, also has this
symmetry to ${\cal{O}}(1/\Lambda^4)$, up to null-operators. 
Thus the interaction lagrangian can be viewed
as a power series in $1/\Lambda^2$.
The symmetry involves interplay
of the Higgs quartic interaction with the Higgs-Yukawa interaction
and implies the relationship, $\lambda = \half g^2$ between the
top--Yukawa coupling, $g$, and Higgs quartic coupling,
$\lambda$, at a high energy scale $ \Lambda \gta$ few TeV. 
We interpret this to be a new physics scale. The top
quark is massless in the symmetric phase, satisfying
the Nambu-Goldstone theorem.
The fermionic shift part of the current is $\propto (1-H^\dagger H/v^2)$,
owing to the interplay of $\lambda$ and $g$,
and vanishes in the broken phase. Hence the Nambu-Goldstone
theorem is trivially evaded in the broken phase 
and the top quark becomes heavy (it is not a Goldstino).
We have $m_t=m_h$,
subject to radiative corrections that can in principle
pull the Higgs into concordance with experiment. \\
\\ \\

\noindent
Invited Plenary Talk at SCGT12, ``KMI-GCOE Workshop on Strong Coupling\\ 
Gauge Theories in the LHC Perspective'', 4-7 Dec. 2012, Nagoya University
\end{abstract}


\maketitle

\section{Introduction}\label{aba:sec1}

The Higgs boson can be viewed as a
``pseudo-dilaton'' in a particular limit \cite{Kitano}. 
There is, of course,  a fundamental distinction between a ``scale invariant Higgs field''
and a ``dilatonic Higgs field.'' 
A scale invariant Higgs field has a vanishing mass term, but can
have a nonvanishing gauge, quartic and Yukawa couplings.
To qualify as a (pseudo) dilatonic Higgs boson, the Higgs potential must be (approximately) flat.  

Consider the pure Higgs Lagrangian (no gauge fields):
\beq
{\cal{L}}_0 =  \partial_\mu H^\dagger \partial^\mu H  - \frac{\lambda}{2}(H^\dagger H- v^2)^2
\eeq
As usual, the groundstate has a minimum for:
\beq
\VEV{H^i} = \theta^i , \qquad \makebox{where we can choose:} \qquad \theta^i = (v, 0),
\eeq
where $\theta^i$ is an arbitrary orientation in
gauge space and can be rotated  under $SU(2)_L\times U(1)$.

In the limit of small
$\lambda $, the Higgs potential plays the analogue role of 
an ``applied external magnetic field'' to a spin system, 
pulling $\VEV{H^i}$ to the minimum VEV,  $v$.
If we then take $\lambda\rightarrow 0$ the Lagrangian 
acquires a global ``shift symmetry,"
\beq
\label{five10}
\delta H^i = \theta^i \epsilon \qquad \longrightarrow \qquad \delta \; \partial_\mu H^\dagger \partial^\mu H = 0
\eeq
The alignment, $\theta^i$, is held fixed 
and the shift is parameterized by the variable $\epsilon$.
The Noether current is then:
\beq
J_\mu = \frac{\delta {\cal{L}}_0}{\delta \partial^\mu \epsilon}
= \theta^\dagger \partial_\mu H + H^\dagger \partial_\mu \theta
\eeq
We see that $\theta$ is a defining part of the current. 
If we view $\theta$ as co-rotating with $H$ under the global $SU(2)\times U(1)$
transformations, the charge $\int d^3x\; J_0$ then commutes with the 
gauge group.

In the broken phase of the theory the current looks more ``dilatonic'':
\beq
 J_\mu \rightarrow \sqrt{2}v\partial_\mu h
\eeq
The dilatonic nature of the Higgs implies that fields that acquire
masses proportional to $v$ are ``scale invariant'' in the sense of spontaneous
scale breaking. That is, we can perform an ordinary scale transformation
which would normally shift mass terms, but we can then undo this  by a compensating shift in $h$. To see this, consider the top quark mass term:
\beq
g\bar{\psi}_L{t_R} H +h.c. \;\; \longrightarrow \;\; m_t\bar{t}t\left(1 + \frac{h}{\sqrt{2}v}\right)
\eeq
Under an ordinary infinitesimal 
scale transformation we have $t(x) \rightarrow (1-\epsilon)^{3/2}
t(x')$ and $h(x) \rightarrow (1-\epsilon)h(x') $ where $x_\mu=(1+\epsilon)x'{}_\mu$, $d^4x = (1+\epsilon)^4d^4x'$.  Hence the action transforms as:
\bea
\int d^4x\; m_t\bar{t}t(x) \left(1 + \frac{h(x)}{\sqrt{2}v}\right) \rightarrow
& \rightarrow & \int d^4x'\left((1+\epsilon) m_t\bar{t}t(x') + m_t\bar{t}t(x')\frac{h(x')}{\sqrt{2}v}\right) 
\eea
The latter expression exhibits the fact that under
ordinary scale transformations the $d=4$ Higgs-Yukawa
interaction is scale invariant, while the $d=3$ mass term 
is  not invariant.

However, with the dilatonic shift symmetry
we can compensate the rescaled mass term by a shift in the Higgs field:
\beq
h(x')  \rightarrow  h(x') - \sqrt{2}v \epsilon
\eeq
and we see that:
\beq
\int d^4x'\left((1+\epsilon) m_t\bar{t}t(x') + m_t\bar{t}t(x')\frac{h(x')}{\sqrt{2}v}\right) 
\rightarrow 
\int d^4x' \left( m_t\bar{t}t(x') + m_t\bar{t}t(x')\frac{h(x')}{\sqrt{2}v}\right)
\eeq
Hence, the simultaneous application of
the
scale transformation and Higgs shift symmetry
allows us to maintain the symmetry
of the top quark mass term.  The scale symmetry can be
viewed as spontaneously broken with the Higgs boson playing
the role of the Nambu-Goldstone mode. The same 
invariance applies to the  gauge fields, $W$ and $Z$.   Higgs self-interactions that
involve nonzero $\lambda$ would not be invariant under scale transformations
with dilatonic shifts in h. The symmetry is also broken by
scale anomalies (running couplings). 

\section{Generalize to a ``Super-Dilatation''}

The Higgs boson is thus a (pseudo) dilaton if
the shift transformation is a (approximate) symmetry of the action.
Fundamentally it stems
from the exact shift or modular symmetry of the gaugeless Higgs kinetic term,
as in eq.(\ref{five10}):
\beq
\label{five20}
\delta H^i = \theta^i \epsilon 
\eeq
A key point we wish
to emphasize is that $\epsilon$ is the infinitesimal parameter of the transformation, while the orientation, $\theta^i$, is  held fixed.
$\theta^i$ defines a ``ray'' and the shift moves the field along 
this direction in field space.
We take $\theta^i $ to have the same mass dimension
as the Higgs, \ie, dimensions of mass and it is a normalized isospinor,
$\theta^\dagger \theta = v^2$, where we conventionally choose the alignment 
$\theta^i = (v, 0)$.     
Eq.(\ref{five20}) is a symmetry
of the gaugeless Higgs boson kinetic terms,  $\partial H^\dagger \partial H$.
In such a theory the shift symmetry is exact.

We now propose
a generalization of dilatation symmetry for the Higgs boson
that involves a ``super"-symmetric 
relationship between the top and Higgs fields \cite{hill}.  The shift
in the Higgs boson field is now promoted to an operator.
This symmetry is exact in the 
top, with bottom-left, and Higgs, kinetic terms in the gaugeless limit
(up to total divergences).
Consider the top and Higgs
kinetic terms of the standard model with {\it gauge fields set to zero}:
\beq
\label{kt}
 {\cal{L}}_{K} = \bar{\psi}_L i\slash{\partial}\psi_L + \bar{t}_R i\slash{\partial} t_R +\partial H^\dagger \partial H 
\eeq
We define the infinitesimal transformation:
 \bea
\label{trans0}
\delta \psi^{ia}_L = \theta^{ia}_L\eta \epsilon - i\frac{\slash{\partial}H^i\theta^a_R}{\Lambda^2}\epsilon ; &&
\qquad  
\delta \bar{\psi}_{L\;ia} = \bar{\theta}_{L\;ia}\eta \epsilon + 
i\frac{\bar{\theta}_{Ra}\slash{\partial}H_i^\dagger}{\Lambda^2}\epsilon  ;
\nonumber  \\
\delta t^a_R = \theta^a_R\eta \epsilon 
-i \frac{\slash{\partial}H_i^\dagger\theta^{ia}_L}{\Lambda^2}\epsilon ;  &&
\qquad  
\delta \bar{t}_{Ra} = \bar{\theta}_{Ra}\eta \epsilon 
+ i\frac{\bar{\theta}_{Lia}\slash{\partial}H^i}{\Lambda^2}\epsilon  ;
\nonumber  \\
\delta H^i = \frac{\bar{\theta}_{Ra}\psi^{ia}_L + \bar{t}_{Ra}\theta^{ia}_L}{\Lambda^2}\epsilon ; 
&&
\qquad  \delta H_i^\dagger = \frac{\bar{\psi}_{Lai}\theta^a_R + \bar{\theta}_{Lai}t^a_R}{\Lambda^2}\epsilon .
\eea
where $i$ ($a$) is an 
isospin (color) index. 
$\eta$ is a relative normalization factor that we determine
subsequently.

It is readily seen that eq.(\ref{trans0}) is an invariance of eq.(\ref{kt})
up to total derivatives:
\bea
\label{transkt}
\delta (\bar{\psi}_Li\slash{\partial}\psi_L) & = & \frac{(\bar{\psi}_L\theta_R)\cdot \partial^2 H}{\Lambda^2}\epsilon + h.c.+t.d.
\nonumber  \\
\delta (\bar{t}_Ri\slash{\partial}t_R) & = & \frac{ (\bar{\theta}_L t_R)\cdot \partial^2 H}{\Lambda^2}\epsilon + h.c.+t.d.
\nonumber  \\
\delta (\partial H^\dagger \partial H) & = & -\frac{(\bar{\psi}_L\theta_R + \bar{\theta}_Lt_R)\cdot\partial^2H}{\Lambda^2}\epsilon
+ h.c. + t.d.
\nonumber \\
\makebox{hence,} & &  \delta  {\cal{L}}_{K}  =  0 + t.d.
\eea
The symmetry of the gauge free kinetic terms 
makes no use of equations of motion or on-shell conditions.
At this stage,  the shifts 
in $\psi_L$ and $t_R$ by $\theta_{L,R}$ proportional to $\eta$
play no role, but
will be essential with the Higgs-Yukawa interaction and Higgs mass term. Indeed, shifting $\bar{\psi}_Li\slash{\partial}\psi_L
\rightarrow \bar{\psi}_Li\slash{\partial}\theta_L$ yields a total divergence
provided we have switched off the local gauge fields.  The fermionic shift symmetry,
however, raises 
an issue of consistency of a massive top quark with the Nambu-Goldstone theorem which has a
remarkable resolution, as shown in Section 3.

This transformation exploits the interplay of the quantum 
numbers of $\psi_L$, $t_R$ and $H$. It resembles a scalar supermultiplet 
transformation of component fields, where 
the Higgs field is treated as a superpartner
of $\psi_L$.\cite{Wess}  We emphasize that this is not a representation of the supersymmetry algebra, as there is no ``F'' auxillary field.\cite{Wess} 
This is essentially  a scalar supermultiplet transformation 
with fixed $F=0$ and the superparameters replaced by $\theta\epsilon$).
With the assignment of scales of the $\theta_{L,R}$
and the presence of $\Lambda$ the commutators of subsequent transformations for
different $\theta_{L,R}$ cannot close. Also, the $\theta_{L,R}$ carry
flavor and color quantum numbers, and the failure of the algebra to close
into a superalgebra is presumably a supersymmetric extension of the Coleman-Mandula no-go theorem.   In fact, this is a $U(1)$ symmetry with the transformation parameter, $\epsilon$, 
for fixed background values of $\theta_{L,R}$. As such, the commutator 
trivially vanishes on the fields:
$
 [\delta_{\epsilon{}'}, \delta_{\epsilon}](\psi, H, t_R) = 0
$

We presently turn to the full Lagrangian
of the top-Higgs system in the standard model with gauge fields turned off:
\bea
\label{full}
{\cal{L}}_H & = &  i\bar{\psi}_L \slash{\partial}\psi_L + i\bar{t}_R \slash{\partial} t_R +\partial H^\dagger \partial H 
\nonumber \\
& & + g(\bar{\psi}_L t_R H + h.c.)
- M_H^2 H^\dagger H - \frac{\lambda}{2} (H^\dagger H)^2
\eea
From eq.(\ref{trans0}) we compute the transformations:
\bea
\label{trans2}
\delta (-M_H^2 H^\dagger H) & = & 
-\frac{\epsilon}{\Lambda^2} \; M_H^2(\bar{\psi}_{L}\theta_R + \bar{\theta}_{L}t_R)\cdot H + h.c
 \\
 \label{transquartic}
\delta (-\frac{\lambda}{2}(H^\dagger H)^2) & = & -\frac{\epsilon}{\Lambda^2}\; \lambda(\bar{\psi}_{L}\theta_R + \bar{\theta}_{L}t_R)\cdot H H^\dagger H  + h.c.
\\
\label{transYuk}
\delta (g\bar{\psi}_L t_R H + h.c.) & = & 
g\eta\epsilon(\bar{\psi}_L\theta_R  
+ \bar{\theta}_L t_R) H  
+g^2\frac{\epsilon}{2\Lambda^2} (\bar{\theta}_R \psi_L
+ \bar{t}_{R} \theta_L)\cdot \left( H^\dagger H^\dagger H \right) 
 \nonumber \\
& & + \; g\frac{\epsilon}{\Lambda^2}\bar{\psi}_L t_R(\bar{\theta}_{R}\psi_L + \bar{t}_{R}\theta_L)
\nonumber \\
& & +ig\frac{2\epsilon}{\Lambda^2}\bar{\psi}_L\gamma_\mu\frac{\tau^A}{2}\theta_L 
\left( H^\dagger \stackrel{\leftrightarrow}{\partial^\mu} \frac{\tau^A}{2} H \right)
+ig\frac{\epsilon}{2\Lambda^2} \bar{\psi}_L\gamma_\mu\theta_L 
\left( H^\dagger \stackrel{\leftrightarrow}{\partial^\mu} H \right)
\nonumber \\
& &
-ig\frac{\epsilon}{\Lambda^2}\bar{\theta}_{R}\gamma_\mu t_R 
\left( H^\dagger \stackrel{\leftrightarrow}{\partial^\mu} H \right)  + h.c. + t.d.
\eea
where we use the isospin Fierz identity,  $[\tau^A]_{ij} [\tau^A]_{kl} = 2\delta_{il}\delta_{kj} -\delta_{ij}\delta_{kl}$, 
and: $\stackrel{\leftrightarrow}{\partial^\mu}=\half(\stackrel{\rightarrow}{\partial^\mu}
-\stackrel{\leftarrow}{\partial^\mu})$.
Here we have applied {\it the fermionic equations of motion}:
\beq
i\slash{\partial}t_R + g\psi_L\cdot H^\dagger = 0
\qquad \qquad
i\slash{\partial} {\psi}_L + gt_R H =0
\eeq 
and eq.(\ref{transYuk}) follows \cite{hill}.

Notice in eq.(\ref{transYuk}) we have 
generated a set of higher dimension operator
terms of the form:
\bea
& & \frac{g\epsilon}{\Lambda^2}\bar{\psi}_L t_R(\bar{\theta}_{R}\psi_L  +  \bar{t}_{R}\theta_L)
 +\; i\frac{2g\epsilon}{\Lambda^2}\bar{\psi}_L\gamma_\mu\frac{\tau^A}{2}\theta_L 
 (H^\dagger \stackrel{\leftrightarrow}{\partial^\mu} \frac{\tau^A}{2} H )
\nonumber \\
& &
 +i\frac{g\epsilon}{2\Lambda^2}\bar{\psi}_L\gamma_\mu\theta_L 
 ( H^\dagger \stackrel{\leftrightarrow}{\partial^\mu} H )
-\; i\frac{g\epsilon}{\Lambda^2} \bar{\theta}_{R}\gamma_\mu t_R 
( H^\dagger \stackrel{\leftrightarrow}{\partial^\mu} H )  + h.c. + t.d.
\eea
In analogy to a ``bottoms up'' derivation
of a nonlinear chiral Lagrangian (see section (3) for a review), 
these terms can be cancelled by adding
higher dimension operators to the original
Lagrangian of the form:
\bea
\label{D6}
 {\cal{L}}_{d=6} & = &  \frac{\kappa}{\Lambda^2}(\bar{\psi}_L t_R\bar{t}_{R}\psi_L )
+\frac{2\kappa}{\Lambda^2}(\bar{\psi}_L\gamma_\mu\frac{\tau^A}{2}\psi_L )
(H^\dagger i\stackrel{\leftrightarrow}{\partial^\mu} \frac{\tau^A}{2} H )  
\nonumber \\
& &
+\frac{\kappa}{2\Lambda^2} (\bar{\psi}_L\gamma_\mu\psi_L)
( H^\dagger i\stackrel{\leftrightarrow}{\partial^\mu} H ) 
- \frac{\kappa}{\Lambda^2} (\bar{t}_{R}\gamma_\mu t_R) 
( H^\dagger i\stackrel{\leftrightarrow}{\partial^\mu} H )  
\eea
where we will presently relate the coupling constant, 
$\kappa$, to $M_H$, $\Lambda$ and $g$ below.

We thus obtain the effective Lagrangian, 
\bea
\label{full0}
{\cal{L}}_H & = &  \bar{\psi}_L i\slash{\partial}\psi_L + i\bar{t}_R \slash{\partial} t_R +\partial H^\dagger \partial H 
\nonumber \\
& & + g(\bar{\psi}_L t_R H + h.c.)
- M_H^2 H^\dagger H - \frac{\lambda}{2} (H^\dagger H)^2 
\nonumber \\
& & +\frac{\kappa}{\Lambda^2}(\bar{\psi}_L t_R\bar{t}_{R}\psi_L )
+\frac{2\kappa}{\Lambda^2}(\bar{\psi}_L\gamma_\mu\frac{\tau^A}{2}\psi_L )
(H^\dagger i\stackrel{\leftrightarrow}{\partial^\mu} \frac{\tau^A}{2} H )  
\nonumber \\
& &
+\frac{\kappa}{2\Lambda^2} (\bar{\psi}_L\gamma_\mu\psi_L)
( H^\dagger i\stackrel{\leftrightarrow}{\partial^\mu} H ) 
- \frac{\kappa}{\Lambda^2} (\bar{t}_{R}\gamma_\mu t_R) 
( H^\dagger i\stackrel{\leftrightarrow}{\partial^\mu} H )  
\eea
Performing the super-dilatation
transformation of eq.(\ref{trans0}) we now demand that:
\beq
\delta {\cal{L}}_H  = 0 + {\cal{O}}\left(\frac{1}{\Lambda^4} \right)
\eeq
The generated ${\cal{O}}(1/\Lambda^4) $ terms
can be compensated by adding additional $1/\Lambda^4$ terms to
the Lagrangian. By continued iteration of eq.(\ref{trans0}) we 
would generate a power series  of contact interactions 
that are scaled by $\sim 1/\Lambda^{2n}$.  

 First we see that the transformation of the Higgs mass term of eq.(\ref{full0}), 
from eqs.(\ref{trans2}--\ref{transYuk}),   
cancels against the first term of the transformed Higgs-Yukawa interaction, provided:
\beq
\label{21}
 g\eta  = \frac{M_H^2 }{\Lambda^2} 
\eeq
(beware: $M_H^2$ is the
negative Lagrangian (mass)$^2$, while $m_h^2 = -2M_H^2$ is the physical 
positive Higgs boson (mass)$^2$ in the broken phase,
and we take the normalization $v^2 = \VEV{H^\dagger H} =(175\; GeV)^2$ )
This establishes
the normalization factor, $\eta$.
It also establishes the relative sign (we assume $\Lambda^2$ positive). 
If we're in the symmetric (broken) phase, 
$M_H^2$ positive (negative), then we have $g\eta >0$ ( $g\eta <0$).
We have the freedom of choosing arbitrary $\eta$ 
since the defining kinetic term invariance involves only $\epsilon$.

One might think we can now take $\Lambda^2$ to be arbitrarily large compared
to $M_H^2$ by adjusting $|\eta| << 1$. However,  the
second term of eq.(\ref{transYuk})
must also cancel against the transformation of the first $\kappa$ term appearing in eq.(\ref{full0}). This requires that:
\beq
\label{above}
\eta{\kappa}= - g,
\qquad \makebox{or, using eq.(\ref{21}):} \qquad \frac{\kappa}{\Lambda^2} = -\frac{g^2}{M_H^2}
\eeq
This is  a striking result: a seesaw relation between the
weak scale and $\Lambda$-scale terms. In the $d=6$ operators
we have a Nambu-Jona-Lasionio component. The Nambu-Jona-Lasinio attractive interaction corresponds to $\kappa >0$ , and we see
in eq.(\ref{above}) that the super-dilatation is then 
consistent only if $M_H^2 < 0$.  
Moreover, to make $\eta$ small requires  taking $\kappa$ large.

Finally, the most
interesting relationship, which is the analogue of
the Goldberger-Treiman relationship in a chiral Lagrangian (see section 3),
arises
from the cancellation of the $\sim \epsilon(\bar{\psi}_{L}\theta_R + \bar{\theta}_{L}t_R)\cdot H H^\dagger H$ terms of eqs.(\ref{transquartic}) and (\ref{transYuk})
under the super-dilatation symmetry:
\beq
\label{cc0}
0 = (\lambda - \half g^2)\frac{\epsilon}{\Lambda^2}(\bar{\psi}_{L}\theta_R + \bar{\theta}_{L}t_R)\cdot H H^\dagger H  + h.c
\eeq
or,
\beq
\label{cc}
\lambda = \half g^2 
\eeq
Note that his transformation does
not involve $\eta$.

The $\lambda = g^2/2$ relationship refers to the coefficient of the $D=6$ operator,
$(\bar{\theta}_{L}t_R)\cdot H H^\dagger H +h.c. $    We
therefore assume that it applies at the scale $\Lambda$. 
The low energy relationship between the couplings
$g^2$ and $\lambda$ then
depends upon the  renormalization group running from $\Lambda $ to $v_{weak}
\approx 175$ GeV. 
If we ignore the RG running 
then eq.(\ref{cc}) would hold at the weak scale, and 
implies the supersymmetric relationship
 $m^2_{h} = 2\lambda v^2_{weak}  =m^2_{t}$
in the broken phase.  
This is an improvement over the usual NJL result, $m^2_h = 4m^2_t$
(though we were seeking $m_h^2 = m_t^2/2$).

Note that we can Fierz rearrange the first term of eq.(\ref{D6}):
\beq
  (\bar{\psi}^a_Lt_{Ra})_i(\bar{t}_{Rb}\psi^b)^i
\rightarrow
- (\bar{\psi}_{iL} \gamma_\mu \frac{\lambda^A}{2} \psi^i_L )
(\bar{t}_{R} \gamma^\mu \frac{\lambda^A}{2} t_{R}) + {\cal{O}(1/N)}
\eeq
where $N=3$ is the number of colors.  
This term is a pure Nambu-Jona-Lasinio interaction as arises in topcolor \cite{yamawaki,BHL,topc} in the form of a (color current)$\times$(color current).
Indeed, massive Yang--Mills boson exchange for a boson of mass $M^2$ and
momentum exchange $q^2 < M^2$ produces the negative sign for (current)$\times$(current) interactions.
A positive sign for the first term 
of eq.(\ref{D6}) is the attractive sign for the Nambu-Jona-Lasinio model, 
and we thus see that the
 attractive sign 
corresponds to the correct (negative) sign for 
topgluon exchange.  However, we see that the isospin 
(current)$\times$(current) interaction (second term of eq.(\ref{D6}))
then has the wrong sign (positive) for a gauge boson exchange.
 
Since all of the higher dimension $d=6$ operators  are of the form (current)$\times$(current), they preserve the chirality of the
fermions.  That is, the terms of eq.(\ref{D6}) 
contain
no cross terms of the form $\bar{\psi}_L H t_R(H^\dagger H)^p$.
They thus admit the discrete symmetry, $\psi_L\rightarrow (-1)^N\psi_L$
and $t_R\rightarrow (-1)^{N+1} t_R$.
Operators of mixed chirality can therefore be excluded, or suppressed, on symmetry grounds.
If they are introduced they can be small effects,  controlled by the symmetry limit.

\section{Analogy to a Chiral Lagrangian}

For comparison,  we quickly
review a familiar
derivation of a pion chiral Lagrangian from the ``bottoms-up.''
Consider the kinetic terms:
\beq
{\cal{L}}_K = \bar{\psi}_L i\slash{\partial} \psi_L +  \bar{\psi}_R i\slash{\partial} \psi_R
+ \half \partial_\mu \pi \partial^\mu \pi
\eeq
We'll consider the RH-chiral symmetry:
\bea
\delta \psi_L & = & 0  
\nonumber  \\
\delta \psi_R & = & i\theta\psi_R   
\qquad  
\delta \bar{\psi}_{R} = -i\theta\bar{\psi}_R
\nonumber  \\
\delta \pi & = & f_\pi \theta
\eea
and we demand the Lagrangian is invariant under
this global transformation:
\beq
\delta {\cal{L}}_K  = 0
\eeq
The RH-chiral current is:
\beq
-\frac{\delta {\cal{L}}_K}{\delta \; \partial^\mu \theta} 
= \bar{\psi}_R \gamma_\mu \psi_R -   f_\pi \partial_\mu \pi
\eeq
and we assume 
$f_\pi$ is ``determined from experiment,'' \eg, $\pi \rightarrow \mu \nu$
(of course, in the real world this is the left-handed current).

Consider the interactions consisting of
a massive ``nucleon'' coupled to pion:
\beq
{\cal{L}}_V = M \bar{\psi} \psi -  ig\pi \bar{\psi}\gamma^5 \psi
= M \bar{\psi}_L \psi_R -  ig\pi \bar{\psi}_L \psi_R + h.c.
\eeq
We perform the RH-chiral transformation transformation:
\beq
\delta {\cal{L}}_V = \left(i\theta M  -  igf_\pi \theta  + 
  g\pi f_\pi \theta \right) \bar{\psi}_L \psi_R
+  h.c.
\eeq
so, invariance requires:
\beq
\delta {\cal{L}}_V = 0 \qquad \longrightarrow \qquad  g = \frac{M}{f_\pi}
\eeq
which is the Goldberger-Treiman relation.

However, we must also cancel the
``higher order term'' $ \propto \pi \theta \bar{\psi}_L \psi_R$. 
We thus include an ${\cal{O}}(\pi^2)$ term:
 \beq
{\cal{L}}_V \rightarrow  M \left( 1 - \frac{i\pi}{f_\pi} + c \frac{\pi^2}{f^2_\pi} \right) \bar{\psi}_L \psi_R + h.c.
\eeq
Now:
\beq
\delta {\cal{L}}_V \rightarrow  M\left(i\theta -i\frac{f_\pi\theta  }{f_\pi}
+\frac{\pi\theta  }{f_\pi}
 + 2c \frac{\pi}{f^2_\pi} f_\pi\theta   +    ic \frac{\pi^2}{f^2_\pi} f_\pi\theta
\right)\bar{\psi}_L \psi_R
h.c.
\eeq
so:
\beq
\delta {\cal{L}}_V = 0 \qquad \longrightarrow \qquad  g = \frac{M}{f_\pi}, \qquad c = -\half
\eeq
But, now
we must cancel higher order term $ \propto \pi^2 \theta \bar{\psi}_L \psi_R$
which implies an ${\cal{O}}(\pi^3)$ interaction, and so-forth.

We can sum the resulting power series
and we find, iteratively, the solution:
\beq
{\cal{L}}_V  
= M \bar{\psi}_L U \psi_R + h.c. \qquad \qquad U = \exp(i\pi/f_\pi)
\eeq
whence, the symmetric Lagrangian is:
\beq
{\cal{L}} = \bar{\psi}_L i\slash{\partial} \psi_L +  \bar{\psi}_R i\slash{\partial} \psi_R
+ \frac{f_\pi^2}{2} \partial_\mu U^\dagger \partial^\mu U + M \bar{\psi}_L U \psi_R + h.c.
\eeq
and we have obtained the``nonlinear $\sigma$-model Lagrangian." 

Our present strategy is similar. We begin with the super-dilatational invariance
of the top-Higgs kinetic terms. We then analyze the transformation of the
Higgs-Yukawa, Higgs mass and quartic interactions. We demand overall
invariance of the Lagrangian. We thus find the ``Goldberger-Treiman'' relationship, $\lambda = g^2/2$, which implies $m_t=m_h$
in the broken phase. This induces higher dimension operators. Ultimately,
we expect to sum the tower of operators, though in the present
case we expect that these arise via new dynamics, such as heavy recurrences of
composite Higgs bosons and vector--like top quarks.

\section{Current Structure and the Nambu-Goldstone Theorem}

The critical aspect of our construction is that
the operator shift of $\delta H$ in the quartic Higgs interaction
is cancelling against the super-rotation
(\ie,  the ``twist'') of $\delta \psi$ in the Higgs-Yukawa interaction.
Moreover, the pure fermionic shift in  
$\delta \psi \sim \eta\epsilon \theta $, in the Higgs-Yukawa interaction,
\ie, proportional to $\eta$,
cancels against the $\delta H$ shift in the Higgs mass term.
This ties the transformations together into a single structure.

The Nambu-Goldstone theorem for a
fermionic shift $\delta \psi \sim \eta\epsilon \theta $, which would naively imply a massless top quark (a ``Goldstino''),
is evaded in the broken phase.
How does our theory evade the existence of a zero-mode
associated with the fermionic shift?  Naively, this would
seem to prohibit a massless top quark. 
In fact, this happens in a subtle way. One must carefully
construct the currents given our use of equations of
motion in $\delta (g\bar{\psi}_L t_R H + h.c.)$.
We therefore wish to clarify the 
the relationship to the
Nambu-Goldstone theorem in the present set up.

We consider, for technical simplicity, a simpler 
``minimal'' transformation defined by $\theta_L=0$:
\bea
\label{trans20a}
\delta \psi^{ia}_L =  - i\frac{\slash{\partial}H^i\theta^a_R}{\Lambda^2}\epsilon ; &&
\qquad  
\delta \bar{\psi}_{L\;ia} = 
i\frac{\bar{\theta}_{Ra}\slash{\partial}H_i^\dagger}{\Lambda^2}\epsilon  ;
\\
\label{trans20b}
\delta t^a_R = \theta^a_R\eta \epsilon ;
&&
\qquad  
\delta \bar{t}_{Ra} = \bar{\theta}_{Ra}\eta \epsilon ;
\\
\label{trans20c}
\delta H^i = \frac{\bar{\theta}_{Ra}\psi^{ia}_L }{\Lambda^2}\epsilon ; 
&&
\qquad  \delta H_i^\dagger = \frac{\bar{\psi}_{Lai}\theta^a_R }{\Lambda^2}\epsilon .
\eea
The parameter $\eta $ is still
fixed by the symmetry
interplay of the Higgs mass term and Yukawa interaction
as in eq.(\ref{21}),
\beq
\label{eta20}
g\eta = \frac{ M_H^2}{\Lambda^2} = -\frac{ \lambda v^2}{\Lambda^2}
\eeq
Consider the top and Higgs
system of the standard model with gauge fields set to zero:
\bea
\label{full30}
{\cal{L}}_H & = &  
{\cal{L}}_{K} + g(\bar{\psi}_L t_R H + h.c.)
- M_H^2 H^\dagger H - \frac{\lambda}{2} (H^\dagger H)
\eea
\beq
\label{kt20}
 {\cal{L}}_{K} = \bar{\psi}_L i\slash{\partial}\psi_L + \bar{t}_R i\slash{\partial} t_R +\partial H^\dagger \partial H 
\eeq
It is readily seen that eqs.(\ref{trans20a}--\ref{trans20c}) is a global invariance of eqs.(\ref{kt20})
up to total derivatives. We presently allow $\epsilon$ to be a
local function of spacetime $\epsilon(x)$ (note that the derivatives in   eq.(\ref{trans20a}) act only upon $H$ and not upon $\epsilon(x)$).
We have:
\bea
\label{transkt20}
\delta (\bar{\psi}_Li\slash{\partial}\psi_L) & = & \frac{(\bar{\psi}_L\theta_R)\cdot \partial^2 H}{\Lambda^2}\epsilon + \frac{(\bar{\psi}_L\gamma_\mu 
\slash{\partial}H \theta_R)  }{\Lambda^2} \partial^\mu\epsilon  +  h.c.+t.d.
\nonumber  \\
\delta (\bar{t}_Ri\slash{\partial}t_R) & = & i(\bar{t}_R\slash{\partial}\theta_R)\eta\epsilon + i(\bar{t}_R\gamma_\mu \theta_R)\eta {\partial}^\mu\epsilon +h.c.+t.d.
\nonumber  \\
\delta (\partial H^\dagger \partial H) & = & -\frac{(\bar{\psi}_L\theta_R)\cdot\partial^2H}{\Lambda^2}\epsilon
+\frac{(\bar{\psi}_L\theta_R )\cdot\partial_\mu H}{\Lambda^2}\partial^\mu\epsilon
+ h.c. + t.d.
\eea
The kinetic terms thus lead to a Noether current:
\beq
J^K_\mu = \frac{ \delta {\cal{L}}_{K} }{ \delta  \partial_\mu \epsilon }
=  
i(\bar{t}_R\gamma_\mu \theta_R)\eta+\frac{(\bar{\psi}_L\gamma_\mu 
\slash{\partial}H \theta_R) }{\Lambda^2}
+\frac{(\bar{\psi}_L\theta_R )}{\Lambda^2}\partial_\mu H + h.c.
\eeq
The symmetry of the full action, as 
we have emphasized, involves a cancellation
of the shift of eqs.(\ref{trans20c})  in the Higgs quartic
interaction against the ``twist'' of eq.(\ref{trans20a}) in the Higgs-Yukawa
interaction.  In calculating the 
transformation of the Higgs-Yukawa interaction, however, 
we make use of  an ``integration by
parts'' and discard total divergences
(and subsequently use the fermion equations of motion).  
This integration by parts in the ``twist'' of eq.(\ref{trans20a})
causes the derivative to act upon the  parameter $\epsilon(x)$:
\bea
\label{trans220}
\delta (-M_H^2 H^\dagger H) & = & 
-\frac{\epsilon}{\Lambda^2} \; M_H^2(\bar{\psi}_{L}\theta_R )\cdot H + h.c
 \\
\delta (-\frac{\lambda}{2}(H^\dagger H)^2) & = & -\frac{\epsilon}{\Lambda^2}\; \lambda(\bar{\psi}_{L}\theta_R )\cdot H H^\dagger H  + h.c.
\\
\label{transYuk20}
\delta (g\bar{\psi}_L t_R H + h.c.) & = & 
g\eta\epsilon(\bar{\psi}_L\theta_R ) H  
+\frac{g^2\epsilon}{2\Lambda^2} (\bar{\theta}_R \psi_L)\cdot \left( H^\dagger H^\dagger H \right) 
 \nonumber \\
& & + \; \frac{g\epsilon}{\Lambda^2}(\bar{\psi}_L t_R)(\bar{\theta}_{R}\psi_L )
-i\frac{g\epsilon}{\Lambda^2}\bar{\theta}_{R}\gamma_\mu t_R 
\left( H^\dagger \stackrel{\leftrightarrow}{\partial^\mu} H \right)  
\nonumber \\
& & 
-i \frac{g}{2\Lambda^2}\bar{\theta}_{R}\gamma_\mu  t_R 
(H^\dagger H)\partial^\mu\epsilon   + h.c. + t.d.
\eea
The last term in eq.(\ref{transYuk20}) shows explicitly that the result of the
integration by parts leads to an additional term $\propto \partial^\mu\epsilon$.
This, in turn, modifies the current, which now becomes:
\beq
J_\mu = \frac{ \delta {\cal{L}}_{H} }{ \delta  \partial_\mu \epsilon }
=   
i(\bar{t}_R\gamma_\mu \theta_R)\left(\eta + \frac{g H^\dagger H}{2\Lambda^2} \right)
+\frac{(\bar{\psi}_L\gamma_\mu 
\slash{\partial}H \theta_R) }{\Lambda^2}
+\frac{(\bar{\psi}_L\theta_R )}{\Lambda^2}\partial_\mu H +h.c.
\eeq
Using the relationship
of  eq.(\ref{eta20}), $g\eta = -\lambda v^2/\Lambda^2$, and the ``Goldberger-Treiman'' analogue,
$\lambda = g^2/2$, the current
can be written:
\beq
\label{current20}
J_\mu{} = \frac{ \delta {\cal{L}}_{H} }{ \delta  \partial_\mu \epsilon }
=   
i\eta(\bar{t}_R\gamma_\mu \theta_R)\left(1 - \frac{H^\dagger H}{v^2} \right)
+\frac{(\bar{\psi}_L\gamma_\mu 
\slash{\partial}H \theta_R) }{\Lambda^2}
+\frac{(\bar{\psi}_L\theta_R )}{\Lambda^2}\partial_\mu H+h.c.
\eeq
The modification of the current occurs in the first term
which is associated with the fermionic ``shift''-symmetry of $t_R$.
Again, this arises from the crucial linkage of the
$\delta H$ shift in the quartic Higgs interaction to
the $\delta \psi_L$ shift in the Higgs-Yukawa interaction.

Note that, in the broken phase where $\VEV{H}=v$ the modification
of the current has the effect of ``turning off'' the fermionic shift.
Indeed,  we will now see that this has a remarkable effect
in evading the Nambu-Goldstone theorem in the broken phase,
and permitting the top quark to be massive.

Consider the two-point function of our current of eq.(\ref{current20})
with $t_R$:
\beq
S(y) = \int \; d^4x e^{iq\cdot x} \;\partial^\mu \bra{0}T^* J_\mu(x)\;t_R(y)\ket{0}
\eeq
($T^*$ implies anti-commutation in the ordering of fermion fields).
Formally, with $\partial^\mu J_\mu = 0$,
we have, from the $\partial_0$ acting upon the $T^*$ ordering
a $\delta(x^0-y^0)$, and:
\bea
\label{Ward1}
S(y) & = & \int \; d^3x\; e^{iq\cdot x} \bra{0} \{ J_0(x),\;t_R(y)\}\ket{0}
\nonumber \\
& = & \int \; d^3x\; e^{-i\vec{q}\cdot \vec{x}} \bra{0} \{ J_0(\vec{x}),\;t_R(\vec{y})\}_{e.t.}\ket{0}
\nonumber \\
& = &   \bra{0} \{Q,\;t_R(\vec{y})\}\ket{0}
\eea
where the charge operator $Q$ is:
\beq
Q = \int \; d^3x\; J_0(\vec{x}).
\eeq

In the symmetric phase of the standard model we have
the Higgs VEV,  $\VEV{H} =0$, and we can neglect all
terms in the current that involve $H$.
The charge operator then involves only the first term in $J^K_\mu
= i\eta\bar{t}_R\gamma_\mu \theta_R + h.c.$,
whence it generates a shift in the fermion field:
\beq
\label{Ward2}
\bra{0} \{Q,\;t_R(\vec{y})\}\ket{0} = \eta{\theta_R}
\eeq
On the other hand we have:
\bea
\label{Ward3}
S(y) & = & \int \; d^4x \; e^{iq\cdot x} \;\partial^\mu \bra{0}T^* i\bar{t}_R(x)\gamma_\mu \theta_R\eta\; \; \; t_R(y)\ket{0} \nonumber \\
& = & -\int \; d^4x \; e^{iq\cdot x} \;i\partial^\mu \gamma_\mu S_F(x-y)\theta_R\eta  \nonumber \\
& = & \left. \frac{q^2 + \slash{q}m}{q^2-m^2}\theta_R\eta \right|_{q\rightarrow 0}   
\eea
In the $q^2\rightarrow 0$ limit the consistency 
of eq.(\ref{Ward3}) with eqs.(\ref{Ward1}, \ref{Ward2}) requires that 
the fermion mass satisfy $m=0$.
This is the fermionic  Nambu-Goldstone theorem and it informs us that any 
fermionic action which has a pure fermionic shift symmetry, 
must correspond to a massless fermion.
This is, indeed, the case in the symmetric phase 
in which the top quark is massless and $\VEV{H}=0$.

Naively we might conclude that the top quark is forced by our
symmetry to be a Goldstino
and remain massless, even in the broken phase. However, we have seen 
that the current is modified in a significant way
in the present case:
\beq
J_\mu^{{shift}}  =   
i\eta(\bar{t}_R\gamma_\mu \theta_R)\left(1 - \frac{H^\dagger H}{v^2} \right) +h.c.
\eeq
In the broken phase,  when  $\VEV{H}=v\neq 0$, this implies that the pure fermionic shift operator in the current ``turns off:''
\beq
\label{J2zero}
 J_\mu^{{shift}} = 0 
\eeq
This is a consequence of the interplay between the quartic interaction and
the Higgs-Yukawa interaction in our construction. 
It implies that there can exist dynamical situations in which a Goldstino
is massless in a symmetric phase, but acquires mass in a broken
phase of a theory. The underlying fermionic shift, $\delta \psi =\eta \theta$ is intact, but
the current is modified dynamically to evade the naive Nambu-Goldstone
theorem.  One might wonder what happens for $\VEV{H^\dagger H} \neq v^2$ and $\neq 0$?
This is, of course, and unstable vacuum and the S-matrix derivation fails. 
Of course, the above current algebra analysis serves only as a 
consistency check on our original Lagrangian analysis, which showed
that a vacuum with massive top and Higgs, with $m_t = m_h$, exists.

Indeed, the symmetry yields
$\lambda = g^2/2$ in both phases of the Standard
Model.  However, the fermionic shift part of the current
is nontrivially modified by the quartic-Yukawa interplay
and is $\propto (1-H^\dagger H/v^2)$. It thus vanishes in
the broken phase with $\VEV{H}=v$.   Therefore,
the top quark becomes massive in the broken phase in the usual way, with
the relationship $m_h = m_t$. This is consistent with the Nambu-Goldstone theorem that would
otherwise naively force the top quark to be 
a massless Goldstino.   
This relationship $\lambda = g^2/2$ 
is the analogue of a Goldberger-Treiman'' relationship.
It holds at a high scale, $\Lambda$, and
is subject to renormalization group and higher dimension operator 
effects that can bring the physical 
masses into concordance with $m_h^2 \approx m_t^2/2$.

\section{UV-Completion and Conclusions }

The effective Lagrangian we obtain from the minimal
transformation of eq.(\ref{trans20a}) is fairly simple:
\bea
\label{full10}
{\cal{L}}_H & = &  \bar{\psi}_L i\slash{\partial}\psi_L + i\bar{t}_R \slash{\partial} t_R +\partial H^\dagger \partial H 
\nonumber \\
& & + g(\bar{\psi}_L t_R H + h.c.)
- M_H^2 H^\dagger H - \frac{\lambda}{2} (H^\dagger H)^2 
\nonumber \\
& & +\frac{\kappa}{\Lambda^2}(\bar{\psi}_L t_R\bar{t}_{R}\psi_L )
- \frac{\kappa}{\Lambda^2} (\bar{t}_{R}\gamma_\mu t_R) 
( H^\dagger i\stackrel{\leftrightarrow}{\partial^\mu} H )  + {\cal{O}}\left(\frac{1}{\Lambda^2}\right)
\eea 
We retain the relationships of eqs.(\ref{21},\ref{above},\ref{cc}).

Note the structure of the higher dimension operators.  One of these
is a Nambu-Jona-Lasinio four-fermion interaction, as expected in topcolor \cite{topc}.
The other is a current-current interaction of the Higgs with the top quark. 
With $\Lambda$ sufficiently large, $\kappa$ must become large in accord with
the seesaw relationship of eq.(\ref{above}). 

This is suggestive of a dynamics in which a boundstate recurrence of the Higgs
boson, composed of $\bar{t}t$ is generated via the NJL interaction
\cite{topc,yamawaki,BHL}.  The $d=6$ Higgs-top interactions can likewise generate a composite
Dirac (vectorlike) fermion, composed of $\sim H^\dagger t_R$  with gauge quantum numbers 
of the LH top quark. \cite{bdobhill}   The Dirac fermion has a RH component that is an s-wave, $\sim H^\dagger t_R$, and a LH component that is a p-wave,
$\sim H^\dagger i\slash{\partial} t_R/\Lambda $. 

We believe the tower of operators $\sim 1/\Lambda^{2n}$ may be determined,
and the dynamical model admitting our super-dilatation may be understood
as a full solution to this dynamics.  This may be facilitated by
introducing the composite fields explicitly as auxilliary fields.
These and related issues are under further investigation.\cite{bdobhill}

While the usual superalgebra of SUSY does not permit
the Higgs to be the superpartner of the $(t,b)$ quarks,
the symmetry we present here does accomplish this.
We emphasize that the ``super''-dilatation symmetry is not
a conventional super--algebra, \ie, it is not a grading of the Lorentz Group, and is not associated with a nontrivial nonabelian closed superalgebra
(at least not in our present exploratory formulation). 
The symmetry is a generalization of a dilatonic
shift symmetry for the Higgs, in which the shift
is promoted to an operator, and 
the transformation contains a single 
bosonic parameter $\epsilon$, and we have
 a $\sim U(1)$, invariance, which closes trivially.  

Our symmetry is 
remniscent of a ``reparameterization invariance," \eg, as occurs
in heavy quark effective field theory (HQET) \cite{Georgi,LukeManohar,Heinonen}.  
In the latter case 
one considers an $M\rightarrow \infty$ limit for a heavy quark and constructs a  field theoretic Lagrangian 
for a given four-velocity ``supersector,'' $v_\mu$. The Lagrangian
takes the form of a series expansion in higher dimension operators weighted by powers of $1/M$.  The leading terms in the theory
display heavy-spin symmetry (\eg, degenerate  $0^-$ and $1^{-}$ mesons). 
The reparameterization invariance is a residual symmetry that constrains 
the full operator structure and
relates the coefficients of the terms in the 
Lagrangian to higher orders of $1/M$.  The
reparameterization invariance is essentially the vestige
of the underlying hidden Lorentz invariance \cite{Heinonen}.


\end{document}